\documentclass[preprint,prb,aps,footinbib,amsmath,amssymb,superscriptaddress]{revtex4-1}
\usepackage{graphicx}
\usepackage{dcolumn}
\usepackage{color}
\usepackage{bm}
\usepackage{natbib,hyperref}
\usepackage{amsmath,amssymb,amsfonts,bbold}
\usepackage[normalem]{ulem}
\usepackage{svg}
\usepackage{listings}
\usepackage[caption=false]{subfig}
\usepackage[usestackEOL]{stackengine}
\usepackage{soul}

\newlength{\subcolumnwidth}

\newcommand{\nextsubcolumn}[1][]{%
  \cr\noalign{\hfill}
  \if\relax\detokenize{#1}\relax\else\hsize=#1\setlength{\subcolumnwidth}{\hsize}\fi
}

\newcommand{\ourtitle}{Electron spin resonance with scanning tunneling microscopy: a tool for an on-surface quantum platform of identical qubits}
\begin{document}

\title{\ourtitle}

\author{Deung-Jang Choi}
\email{djchoi@dipc.org}
\affiliation{Centro de F{\'{i}}sica de Materiales
        CFM/MPC (CSIC-UPV/EHU),  20018 Donostia-San Sebasti\'an, Spain}
\affiliation{Donostia International Physics Center (DIPC),  20018 Donostia-San
 Sebasti\'an, Spain}
 \affiliation{Ikerbasque, Basque Foundation for Science, 48013 Bilbao, Spain}
 \author{Soo-hyon Phark}
\affiliation{%
Center for Quantum Nanoscience (QNS), Institute for Basic Science (IBS), Seoul 03760, Republic of Korea
}%
\affiliation{%
Department of Physics, Ewha Womans University, Seoul 03760, Republic of Korea
}%
\author{Andreas J. Heinrich}
\affiliation{%
Center for Quantum Nanoscience (QNS), Institute for Basic Science (IBS), Seoul 03760, Republic of Korea
}%
\affiliation{%
Department of Physics, Ewha Womans University, Seoul 03760, Republic of Korea
}%
 \author{Nicol{\'a}s Lorente}
\affiliation{Centro de F{\'{i}}sica de Materiales
        CFM/MPC (CSIC-UPV/EHU),  20018 Donostia-San Sebasti\'an, Spain}
\affiliation{Donostia International Physics Center (DIPC),  20018 Donostia-San Sebasti\'an, Spain}


\begin{abstract}
Integration of electron spin resonance (ESR) in a scanning tunneling microscope (STM) has enabled an all-electrical control of atomic and molecular spins on solid surfaces with atomic-scale precision and energy resolution beyond thermal limitations. Further, coherent manipulation and detection of individual spins in an ESR-STM establishes a powerful quantum platform, allowing for the implementation of fundamental quantum logic operations to on-surface identical qubits. In this review, we introduce recent advances of ESR-STM, focusing on its application to atomic-scale qubits and extension to molecular qubit systems. We discuss the principles underlying ESR-STM, followed by single-spin addressability, coherent control via Rabi oscillations, and quantum state readout through frequency-resolved detection. We further demonstrate multi-qubit control architectures enabled by atom manipulation and local magnetic field engineering, culminating in the realization of multi-qubit logic gates such as the Controlled-NOT and Toffoli gates. These implementations highlight the specialty of ESR-STM towards atomic-scale quantum circuits. Indeed, ESR-STM can be an excellent tool to perform and evaluate quantum operations in molecular qubits. The results reviewed in this collection establish ESR-STM as a versatile tool for advancing quantum coherent science at the atomic and molecular level in solid-state environments.
\end{abstract}

\date{\today}

\maketitle

\section{Introduction}

Quantum computing represents a paradigm shift in our computational capabilities, offering the potential to solve problems that remain intractable for classical architectures~\cite{Preskill2018}. At its core, quantum computing exploits qubits—quantum analogs of classical bits that can exist in coherent superpositions—to process information in parallel. Various platforms have been explored for qubit realization, ranging from superconducting circuits~\cite{Kjaergaard2020} and trapped atoms~\cite{Henriet2020} to semiconductor quantum dots~\cite{Veldhorst2017}. Despite noteworthy progress in coherence time and gate fidelity, challenges such as the complexity of cryogenic electronics~\cite{Charbon2021} and fabrication limits persist.

Electron spin resonance (ESR)~\cite{Poole1996} offers the high energy resolution required for precise spin measurements, independent of thermal broadening, while scanning tunneling microscopy (STM) provides atomic-scale spatial resolution~\cite{Wiesendanger1994}. The synergistic combination of these techniques in an ESR-STM thus enables addressing of individual spins with energy resolutions reaching tens of neV in the study of single atoms, atomically controlled structures, and molecules~\cite{Baumann_Paul_science_2015,Choi2017a,Willke_Bae_science_2018,Seifert_2020,zhang_electron_2022,kovarik_electron_2022}. 
Construction and coherent control of nanoscale qubit systems are central to advancing quantum-coherent nanoscience~\cite{Heinrich2021}. In this regards, coherent manipulation of single electron spins via radio-frequency (RF) excitation in an ESR-STM~\cite{Yang2019,Veldman2021} have put a milestone for bottom-up approach to construct qubit architectures utilizing individual atoms on a solid surface. However, achieving the full potential of this approach necessitates scaling beyond the confines of a subnanometer junction to incorporate multiple, addressable qubits~\cite{Science}.

In this context, molecular spins emerge as an attractive alternative to conventional two-level systems. Their rich internal structures not only allow for the possibility of multi-level quantum computation but also provide a versatile platform for tailoring energy level architectures that facilitate robust quantum error correction protocols~\cite{Terhal2015,Campbell2017,Atzori2019}. Moreover, the natural propensity of molecules to self-assemble offers a pathway far more efficient than atoms in fabricating high-density qubit arrays—a crucial attribute for scalable quantum computing architectures~\cite{Leuenberger2001,Cornia2020,Warner2013}. The atomic-scale spatial resolution and high energy discrimination inherent to ESR-STM are therefore ideally suited for investigating and controlling the complex quantum states in molecular systems~\cite{Molmer2020,Ardavan2007}.

This article reviews the unique capabilities of ESR-STM to achieve quantum-coherent manipulation of both atomic~\cite{Science} and molecular~\cite{hill_making_2025} qubits. By harnessing the full quantum state space offered by these systems, we seek to overcome current scalability limitations and pave the way for the realization of high-density, robust quantum computing platforms.

\section{ESR-STM}
\begin{figure}
    \centering
    \includegraphics[width = 0.7\textwidth]{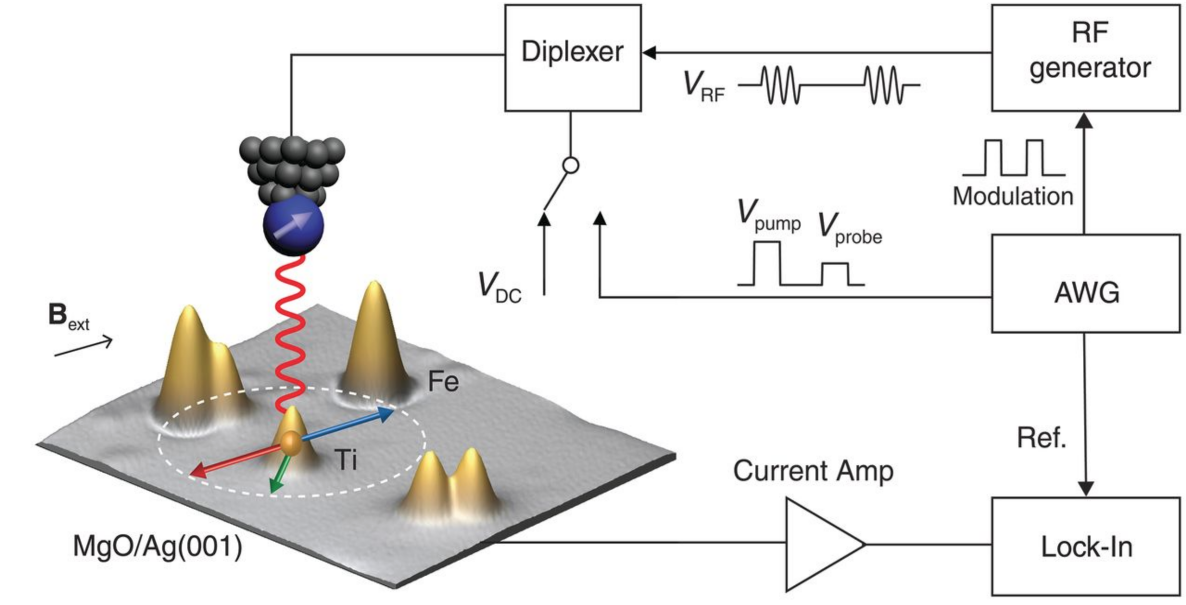}
    \caption{Scheme of a scanning tunneling microscope prepared for performing pulsed ESR experiments on single atoms such as Ti on a MgO/Ag(100) substrate. The STM tip is shown above a surface with several atomic or molecular structures (possible impurities or nanostructures). An radio-frequency (RF) voltage is added via a diplexer on the DC bias of the STM.  The introduction of an arbitrary waveform generator (AWG) in the radio frequency generation leads to the creation of control pulses with a time resolution of 1 ns and sub-milliVolt amplitudes. Adapted from Ref.~[\onlinecite{Science}].}
    \label{Fig_1}
\end{figure}

Coherent control of individual qubits is fundamental to any quantum computation scheme. ESR-STM represents a powerful approach for such control at the atomic and molecular scale~\cite{Heinrich2021,Yang2019,Veldman2021}. In 2015, ESR in an STM was first demonstrated for electron spins localized in single atoms on a surface using a continuous-wave (CW) RF signal.~\cite{Baumann_Paul_science_2015} The next major step forward was the demonstration of pulsed driving of ESR in 2019, which enabled quantum state manipulation of electron spin in a titanium (Ti) atom (Fig.~\ref{Fig_1}).~\cite{Yang2019}

In ESR-STM, spin manipulation is achieved by applying a RF modulated bias voltage to the tunnel junction formed between an STM tip and a spin center adsorbed on a substrate. The RF bias induces transitions in the quantum states of the spin when the radio frequency matches the Larmor frequency of the spin, resulting in a detectable change in tunneling current~\cite{Baumann_Paul_science_2015}. This allows for a precise measurement of spin resonance with an energy resolution down to some tens of neV, far exceeding the limit due to thermal broadening. Moreover, by applying RF pulses of controlled duration and amplitude, it is possible to induce coherent rotations of the spin. The phase, frequency, duration and amplitude of the driving pulse determine the quantum state evolution of the spin, enabling a full control of the corresponding qubit vector on the Bloch sphere~\cite{Yang2019,Phark_2023,Wang_2023}.

Figure~\ref{Fig_1} presents a schematic diagram of the ESR-STM setup, used in the first demonstration on coherent control of single atomic spins on a surface.~\cite{Yang2019} The system combines the STM's capability of spatially addressing individual atoms or molecules with driving of ESR in a time-controlled manner, allowing for coherent control of electron spins of single Ti adsorbates on a 2-monolayer-thick (2 ML) MgO decoupling layer prepared on an Ag(100) substrate.~\cite{Yang2019,Phark_2023,Wang_2023}.
In a simplified scheme, the spin-polarized (SP) tip apex plays a crucial role for ESR experiments in an STM in terms of both driving and probing the magnetic resonance of the spin in the tunneling junction: (i)  its spin-polarization couples to the RF electric field in the junction. It induces time-dependent Hamiltonian matrix elements $\langle 1| \hat{H}_{1}(t) |0\rangle$ required for transition between the two states $|0\rangle$ and $|1\rangle$ of the electron spin~\cite{Lado2017,J_Reina_Galvez_2019,Spin_torque-driven_Kovarik_2024}. (ii) In analogy to a model of magnetic tunnel junction, first the SP-tip can be seen as one magnetic electrode. Second, together with the spin-polarization of the target atom, it induces a tunneling magneto-resistance effect. When driving ESR of the atom, this magneto-resistance changes from that of the non-driven state, resulting in a corresponding change in the tunneling current, which is the readout of ESR-STM. 

For the pulsed-ESR, modulated DC voltage in time from an arbitrary waveform generator output is fed into the RF signal source, gating the width and amplitude of the signal out from the RF signal source. This RF pulse, controlled at a nanosecond time scale, drives a fraction of rotation of the target spin within its coherence time, which enables coherent control experiments such as Rabi oscillation, Ramsey fringe, spin-echo, and so on, as demonstrated first by K. Yang $\it{et~al}$. in 2019~\cite{Yang2019}.

\section{Quantum gates in ESR-STM}

Driving ESR of qubits creates a unitary evolution of the qubits, the so called \textit{quantum gates}. By controlling the time duration of the driving pulse, the evolution can be terminated to leave the qubit in a pre-determined quantum state along a Rabi oscillation. By applying a sequence of pulses, furthermore, different unitary evolutions are chained together, leading to a complex coherent manipulation of the qubit state.

\subsection{Single-qubit gate}
\begin{figure}
    \centering
    \includegraphics[width = 0.4\textwidth]{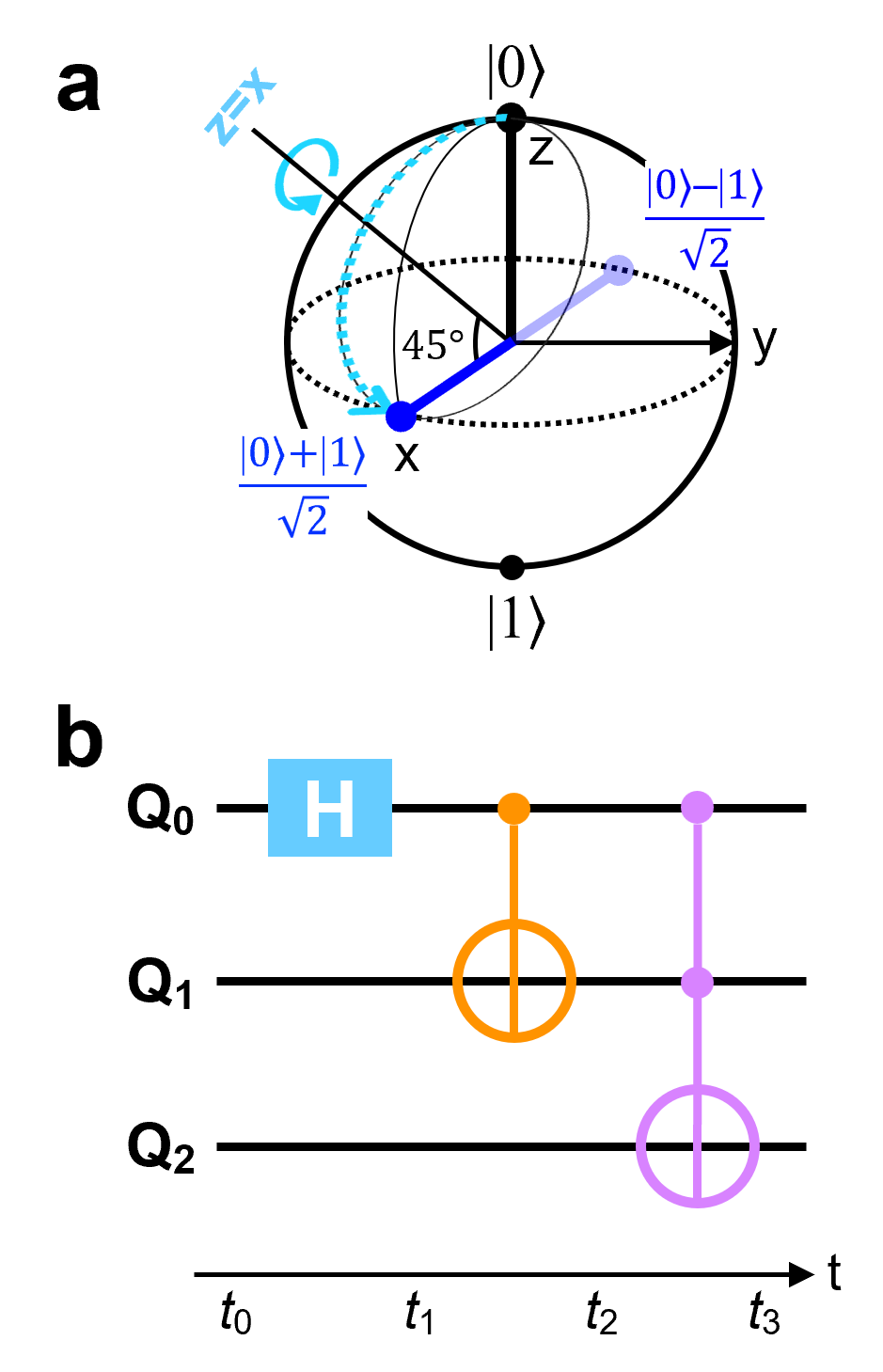}
    \caption{(a) A scheme using the Bloch sphere illustrating $\it{Hadamard}$ gate to a qubit initialized to its $|0\rangle$ state. (b) A Quantum circuit to generate a maximally-entangled \textit{Greenberger-Horne-Zeilinger} (GHZ) state when all three qubits ($Q_0$, $Q_1$, $Q_2$) are initialize to $|0\rangle$ states. The circuit implements a \textit{Toffoli} (\textit{Controlled-controlled-NOT}; CCNOT) gate, the third gate (purple), which flips the state of the third qubit, $Q_2$ if and only if the other two qubits are in the state $|1\rangle$. The first gate (cyan), $\it{Hadamard}$ gate ($H$), creates a superposition state $(|0\rangle + |1\rangle)/\sqrt{2}$ of $Q_0$. The second gate (orange) is a two-qbit gate, \textit{Controlled-NOT} (CNOT), which flips $Q_1$ if and only if $Q_0$ is in the state $|1\rangle$. As a consequence, the consecutive two gates, $\it{Hadamard}$ and CNOT, transform the two-qubits $Q_0$ and $Q_1$ into an entangled state, $(|00\rangle + |11\rangle)/\sqrt{2}$, which finally leads to a GHZ or ``cat'' state, $(|000\rangle + |111\rangle)/\sqrt{2}$, after the \textit{Toffoli} gate.}
    \label{Fig_2}
\end{figure}

A series of RF pulses tuned to the Larmor frequency of a single atom spin can produce desired single-qubit quantum gates in ESR-STM. A simple example is the \textit{Hadamard} gate as illustrated in Fig.~\ref{Fig_2}(a) using the Bloch sphere and vector representation. This gate, when applied to a single spin initialized in its ground state $|0\rangle$, rotates the state vector by an angle of $\pi$ about the axis defined $z = x$, which creates a linear superposition of the two states ($|0\rangle$ and $|1\rangle$) of the qubit, $|+\rangle = (|0\rangle + |1\rangle)/\sqrt{2}$. If the qubit is initialized in its excited state $|1\rangle$, the \textit{Hadamard} gate acts to create another superposition state of the qubit $|-\rangle = (|0\rangle - |1\rangle)/\sqrt{2}$, orthogonal to $|+\rangle$. The gate implemented in this way allows us to retrieve the behavior of the \textit{Hadamard} gate on the selected states.

This set of two operations is very useful when creating an entanglement between two qubits as illustrated in the circuit example of Fig.~\ref{Fig_2}(b). The \textit{Hadamard} gate ($H$ in the cyan box) acting on the first qubit $Q_0$ in its $|0\rangle$ state at $t_0$ transforms $Q_0$ into $|+\rangle$ state at $t_1$, which is going to influence the state of the second qubit $Q_1$ with the two-qubit gate \textit{controlled-NOT} (CNOT; orange) at $t_1$, leading to an entangled state of the two qubits $(|00\rangle + |11\rangle)/\sqrt{2}$ at $t_2$. A straightforward extension of this controlled operation to three-qubit system is the \textit{Controlled-controlled-NOT} (CCNOT; Toffoli) gate, as depicted in purple. Together with the two qubits in the entangled state $(|00\rangle + |11\rangle)/\sqrt{2}$, the CCNOT gate includes the third qubit $Q_2$. The final circuit yields  a fully entangled three-qubit state $(|000\rangle + |111\rangle)/\sqrt{2}$, the so-called \textit{Greenberger–Horne–Zeilinger} (GHZ)~\cite{greenberger_bells_1990} or \textit{cat} state.

\subsection{Two-qubit gate: \textit{Controlled-NOT}}
\begin{figure}
    \centering
    \includegraphics[width = 1\textwidth]{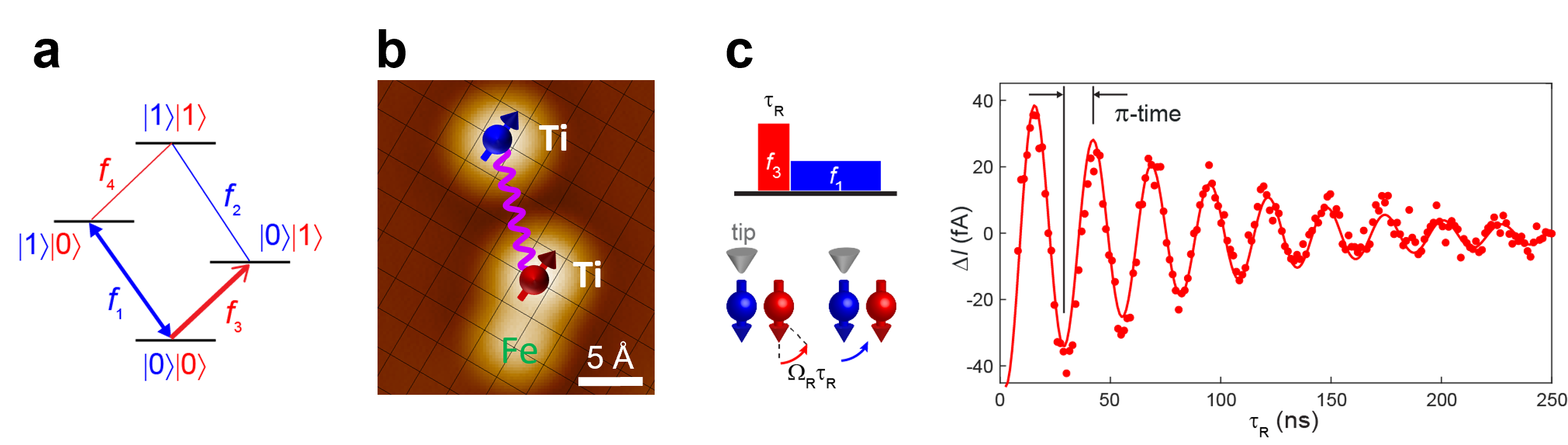}
    \caption{(a) Energy diagram of a two-qubit system, illustrating its four eigenstates $|00\rangle$, $|01\rangle$, $|10\rangle$, $|11\rangle$ and four available quantum transitions represented by four corresponding frequencies $f_1$, $f_2$, $f_3$, $f_4$. (b) A two-qubit structure constructed using two Ti atoms, whose spins are weakly coupled ($\approx110$ MHz) to each other, on a 2-monolayer-thick MgO surface. The ``sensor'' and ``remote'' qubits are depicted with blue and red colors. (c) Schemes of RF pulses and responses of the two Ti spins for a controlled rotation of the ``remote'' qubit (left) and measured Rabi oscillations of the ``remote'' qubit, when the ``sensor'' qubit is in the state $|0\rangle$ (right). This was achieved by a coherent driving of the transition $f_3$ ($|00\rangle \leftrightarrow |01\rangle$) and subsequent steady-state probing using the transition ($|00\rangle \leftrightarrow |10\rangle$) (see the scheme in (a)). The red data points represent the coherent oscillations of the ``remote'' qubit as a function of the driving pulse width $\tau_\mathrm{R}$, characterized by a fit using an exponentially-decaying sinusoidal function. The duration of $\approx13$ ns for the CNOT operation is found as indicated by $\pi$-time.~\cite{Science}}
    \label{Fig_3}
\end{figure}

To be universal, quantum control of multiple qubits only requires a reduced set of single-qubit gates and a two-qubit one. Among numerous two-qubit gates, the CNOT is a typical realization~\cite{Gates} in many available Noisy Intermediate-Scale Quantum (NISQ) architectures~\cite{qiskit2024}. In reality, the CNOT gate can be implemented between two qubits with a minimal interaction obtained by placing them sufficiently far apart on the surface. The weak interaction, in comparison with the detuning, permits to simplify the two-qubit state to products of the two single qubit states (Zeeman product states; $|00\rangle$, $|01\rangle$, $|10\rangle$, $|11\rangle$) with four available quantum transitions, as represented in the energy level diagram of Fig.~\ref{Fig_3}(a). Similar to a single-qubit gate, in such a system we can drive a transition between any adjacent two-qubit states by tuning the RF frequency to the corresponding energy difference ($f_1$, $f_2$, $f_3$, or $f_4$). In Fig.~\ref{Fig_3}(b), a realization of a physical system of such two qubits using two Ti atoms on a 2 ML MgO surface is shown.~\cite{Science}

A CNOT gate is a controlled operation where one  of the qubits (the `control' qubit) selects the action on the other qubit (the `target' qubit). Thus, the NOT gate to the `target' qubit only switches on, by flipping the target's initial state, when the `control' qubit is exclusively in one of its two states, either $|0\rangle$ or $|1\rangle$. 
Here we choose the convention of flipping the second qubit when the first qubit is in the state $|0\rangle$. 
This is achieved by applying a $\pi$-pulse with an RF frequency corresponding to the energy of a specific transition in the two-qubit energy diagram (Fig.~\ref{Fig_3}(a)). 
A $\pi$-pulse is an RF signal applied at the transition frequency for half a Rabi period, driving the system from one state of the transition to the other. 
Figure~\ref{Fig_3}(c) shows a result of Rabi oscillation measurement performed on the transition $|00\rangle \leftrightarrow |01\rangle$ of the two qubits in Fig.~\ref{Fig_3}(b), with the experimental scheme of the pulsed double resonance ESR~\cite{Science}. The flipping of one of the states corresponds to applying a Rabi oscillation by half of its period as indicated by the term `$\pi$-time'. In practice, this inverses the state of the second (red) qubit if and only if the first (blue) qubit is in its $|0\rangle$ state. This RF pulse constitutes a CNOT gate where the `control' qubit is the first and the `target' qubit is the second qubit, Fig.~\ref{Fig_3}.

\subsection{Three-qubit gate: \textit{Toffoli} gate}
\begin{figure}
    \centering
    \includegraphics[width = 1\textwidth]{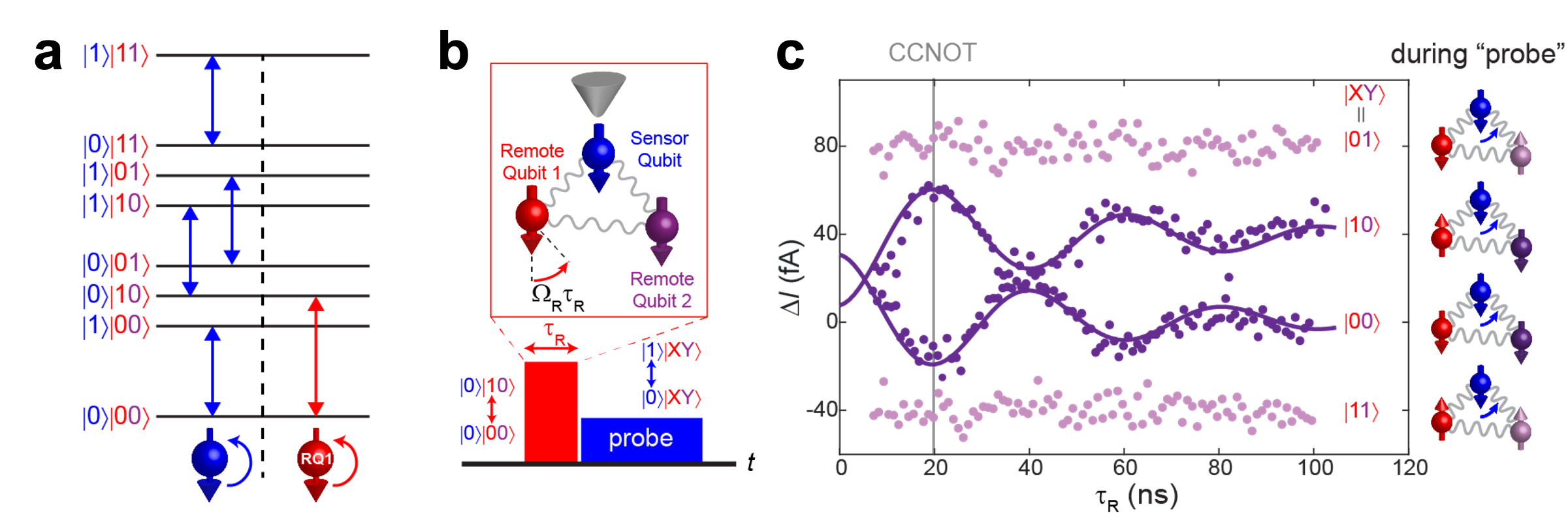}
    \caption{(a) Energy diagram of the eight eigenstates of a three-qubit system, composed of one ``sensor'' (blue) and two ``remote'' (red, purple) qubits. (b) A scheme of pulsed double resonance ESR experiments performed on the three qubits (Ref.~[\onlinecite{Science}]). During the measurement, the tip was parked only on the sensor qubit. The quantum transitions for driving the remote qubit 1 (RQ1; red) and for probing via the sensor qubit (blue) are depicted in (a). Spin arrow pointing down indicates ground state $|0\rangle$ of each qubit. (c) Pulsed double resonance data measured using the probe transitions indicated by the blue arrows in (a). Measurements using two probing transitions for RQ2 in the $|0\rangle$ ($|1\rangle$) state are indicated by dark (light) purple color. Note that this \textit{Toffoli} gate is achieved just by about 20 ns.}
    \label{Fig_4}
\end{figure}

An experimental demonstration of a \textit{Toffoli} (CCNOT) gate to a three-qubit system in ESR-STM has been realized by using three Ti atoms adsorbed onto a 2 ML MgO/Ag(100) surface (Fig.~\ref{Fig_4}). See Ref.~[\onlinecite{Science}] for the details of the physical system. Using the atom manipulation technique, the three atoms are precisely located in a specific environment, allowing for control over their mutual interactions and their coupling to the STM tip and external fields. As a result, the Zeeman product states of the three qubits serve as good approximations to the eigenstates of the system,
preserving the distinct energy levels necessary for the selective addressability of transitions corresponding to multi-qubit logic gates (Fig.~\ref{Fig_4}(a)). RF pulses induce coherent Rabi oscillations between selected quantum states of the three-qubit system. By adjusting pulse amplitude, frequency, and duration, the system undergoes selective transitions corresponding to the logical action of the \textit{Toffoli} gate.

A \textit{Toffoli} gate is carried out by applying a sequence of RF pulses tuned to the transition frequencies between two eigenstates. In Fig.~\ref{Fig_4}(b), a RF pulse scheme for \textit{Toffoli} gate is illustrated for three qubits, composed of one ``sensor'' (blue) and two ``remote'' (red, purple) qubits. Here, the \textit{Toffoli} gate involves flipping the state of the remote qubit 1 (RQ1; red) only when the other two qubits, sensor and remote qubit 2 (RQ2; purple), are in their $|0\rangle$ states. This operation corresponds to the transition between $|000\rangle$ and $|010\rangle$. 

Steady-state driving of four transitions (blue arrows in Fig.~\ref{Fig_4}(a)) are used to probe the controlled rotation of RQ1, and resultant Rabi oscillation data are shown in Fig.~\ref{Fig_4}(c). Coherent rotation of RQ1 appears only from the two probing transitions for RQ2 in the $|0\rangle$ state (dark purple) while no oscillation is observed from the other two for RQ2 in the $|1\rangle$ state (light purple). The Rabi frequencies associated with these transitions are sufficiently large to enable fast gate operations, with state transitions occurring on timescales as short as 20 ns. Such rapid operation times are advantageous, reducing exposure to decoherence and enhancing gate fidelity.

The successful demonstration of a \textit{Toffoli} gate using ESR-STM highlights the potential for atomically precise spin systems to implement complex multi-qubit logic operations beyond single- and two-qubit gates as well as toward more sophisticated quantum operations necessary for fault-tolerant quantum computation. For instance, more complex quantum logic circuits can be realized using atomic spins on a surface, such as the one to create maximally entangled three-qubit states, e.g., the GHZ state that was introduced in Fig.~\ref{Fig_2}, by utilizing the experimental scheme shown in Fig.~\ref{Fig_4}.

\section{Implementation of multi-qubit control in an ESR-STM}
Recent advances have demonstrated that ESR-STM is not limited to single-spin manipulation  as introduced in the section II. It can be extended to the coherent control of multi-qubit systems assembled with atomic precision. However, it necessitates coherent control and measure of spins positioned outside the STM tunnel junction, the so-called ``remote'' qubits, which is implemented via an exchanged interaction with single-atom-magnets (here, Fe atoms; see Fig.~\ref{Fig_3}(b)) that are judiciously placed a few MgO-lattice-constants apart and boost the driving of the ESR signal~\cite{Phark_2023}.
A recent experimental and theoretical studies show that the tip electric field modulates the interaction in time between the Fe and Ti atoms that serves as remote spin, leading to a strong coherent response of the Ti spin~\cite{Phark_2023,Jose}. This approach effectively decouples the control and detection mechanisms of a spin, allowing for scalable multi-qubit architectures. 

Initialization of the spins is achieved thermally by cooling the system to cryogenic temperatures (typically below 1 K) and applying an external magnetic field, resulting in most of the spin population in the ground-state. Detection of remote qubits is performed indirectly through the ESR transition of the ``sensor'' qubit, whose transition frequency depends on the quantum states of the coupled qubits. This transition frequency of the ``sensor'' qubit is carefully chosen by dipolar and exchange interactions with the ``remote'' qubit, which can be engineered through precise control over atomic separations of the two qubits down to the sub-angstrom scale. Multiple remote qubits can be sensed simultaneously and their quantum states can be inferred from the distinct ESR transitions of the sensor qubit. An intriguing extension is to transpose this scheme to multi-magnetic-center molecules where different centers can be assigned to either the sensor or remote qubits, and can be manipulated to achieve multi-qubit quantum operations~\cite{Borilovic2022}.

\subsection*{Molecular Qubits on a surface}
\begin{figure}
    \centering
    \includegraphics[width = 0.8\textwidth]{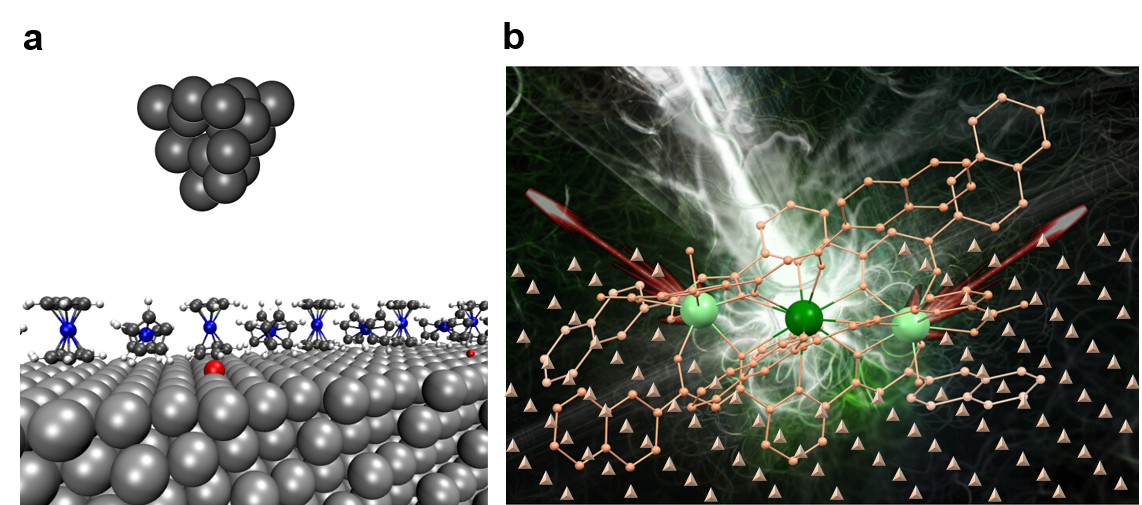}
    \caption{(a) Schematic representation of molecular arrays on a surface, where individual molecules are engineered to address spin centers. These spin centers can be manipulated using Electron Spin Resonance Scanning Tunneling Microscopy (ESR-STM) to perform quantum logic gate operations. This approach enables precise control of quantum states at the atomic level.
(b) Illustration of molecular magnets containing multiple spin centers, which can interact and potentially serve as multi-qubit systems (Adapted from Ref.~[\onlinecite{Aromi2023}]). These molecular architectures offer a promising platform for next-generation quantum computing applications due to their ability to maintain coherence and facilitate spin-based quantum information processing.}
    \label{Fig_5}
\end{figure}

Building these atomic-scale implementations on a solid surface, molecular qubits offer an opportunity to scale the ESR-STM approach to more complex and potentially more robust quantum systems. Molecular complexes can be chemically engineered to host multiple spin centers with tailored coupling strengths and coherence properties~\cite{Cornia2020,Warner2013}. Figure~\ref{Fig_5} illustrates how molecular arrays and molecular magnets can serve as platforms for implementing quantum logic operations. Figure~\ref{Fig_5}(a) shows molecular arrays on a surface that can be addressed individually with the ESR-STM. Figure~\ref{Fig_5}(b) shows a  molecular magnet incorporating multiple spin centers, which demonstrates the potential for multi-qubit interactions. The recent experiments demonstrate that by driving an electron current through one of the active spins in a complex molecular system, other qubits within the molecule can be manipulated via a combination of inter-qubit coupling and time-dependent external fields~\cite{Molmer2020}. Moreover, spectator spins, which are not directly addressed by the RF field, can act as mediators or enhancers of control signals, amplifying the reach of spin manipulation to more remote qubits within the system~\cite{Ardavan2007}.

The above strategy has already been applied to systems of weakly-coupled Ti spins on MgO surfaces. Extending these concepts to chemically designed molecular qubits holds a great promise for realizing universal quantum computation at the molecular level~\cite{Barrios}. By using molecular self-assembly and precision chemistry, it becomes feasible to construct such systems systematically.
Furthermore, the development of multi-qubit gates is of increasing interest due to their ability to reduce the depth of quantum circuits and to implement more efficient error mitigation schemes~\cite{Borilovic2022}. ESR-STM provides the necessary spatial and spectral resolution to implement these gates within molecular systems, offering a route toward scalable, high-density quantum processors.

\section{Conclusions}
Electron spin resonance combined with scanning tunneling microscopy provides a unique and versatile platform for achieving quantum coherent manipulation of atomic and molecular qubits with unparalleled spatial and energy resolution. By combining the local addressability of STM with the spectral selectivity of ESR, ESR-STM offers precise control over individual spins, enabling universal set of quantum gates, composed of single-qubit rotations and CNOT.
In this article, we have summarized the coherent manipulation of multi-qubit systems assembled on insulating surfaces with atomic precision, including the realization of \textit{Controlled-NOT} and \textit{Toffoli} gates. These achievements underscore the potential of ESR-STM for implementing complex quantum logic operations in atomically engineered qubit systems. Furthermore, by extending these techniques to chemically designed molecular qubits, ESR-STM opens new avenues for understanding single-qubit manipulations on the atomic and molecular scales, particularly exploring the extraordinary properties of molecules such as their inherent reproducibility, self-assembly, and tunability.

The successful demonstration of fast and coherent multi-qubit operations at the atomic scale represents an important step toward fault-tolerant quantum computation. The uniqueness of ESR-STM in engineering and addressing spin systems makes it a valuable tool not only for fundamental studies of quantum coherence and entanglement but also for the practical development of high-density quantum processors. Future work will focus on increasing the complexity of qubit architectures, improving coherence times, and integrating molecular qubit networks to advance scalable quantum information processing.

\section*{Data availability}

Figure 1 data for this article are available at DOI: 10.1126/science.ade505.
Figure 3 data for this article are available at DOI: 10.1126/science.ade5050.
Figure 4 data for this article are available at DOI: 10.1126/science.ade5050.
Figure 5(b) data for this article are available at https://doi.org/10.1021/acs.inorgchem.2c03940.

\section*{Acknowledgment}

We thank projects PID2021-127917NB-I00 by MCIN/AEI/10.13039/501100011033, QUAN-000021-01 by Gipuzkoa Provincial Council, PIBA-2024-1-0008 and IT-1527-22 by Basque Government, ESiM project 101046364 by EU. Views and opinions expressed are however those of the author(s) only and do not necessarily reflect those of the EU. Neither the EU nor the granting authority can be held responsible for them. SP and AJH acknowledge supports from Institute for Basic Science (grant: IBS-R027-D1).

\bibliography{references}

\clearpage
\newpage

\end{document}